\documentclass[preprint,prc,showpacs,preprintnumbers,amsmath,amssymb,floatfix]{revtex4}
\usepackage{amsmath}

\usepackage{graphicx}
\usepackage{dcolumn}
\usepackage{bm}
\usepackage{ulem} 
\usepackage[usenames]{color}
\usepackage{epstopdf}
\usepackage{epsfig}
\usepackage{float}
\usepackage{subfigure}
\usepackage{multirow}
\usepackage[version=3]{mhchem}

\newcommand{\nc}{\newcommand}       
\nc{\vc}[1] {\mbox{\boldmath $#1$}} 
\nc{\del}       {\partial}              
\nc{\bra}       {\langle}               
\nc{\ket}       {\rangle}               
\nc{\bras}[1]   {\langle #1|}           
\nc{\kets}[1]   {|#1\rangle}            
\nc{\mapleft}[1]{           
	\smash{\mathop{\,          %
			\hbox to 1.5cm{\rightarrowfill}\, }\limits_{#1}}}
\nc{\beq}     {\begin{eqnarray}} \nc{\eeq}    {\end{eqnarray}}
\nc{\nn}      {\\\nonumber} \nc{\vs}      {\vspace{-0.275cm}}
\nc{\fra}    {\frac{1}{2}}
\nc{\mb}        {\mathbf}

\begin{document}
	
	\preprint{}
	
	\title{The nuclear symmetry energy from relativistic Brueckner-Hartree-Fock model}
	
	\author{Chencan Wang}
	\affiliation{School of Physics, Nankai University, Tianjin 300071,  China}
	\author{Jinniu Hu}
	\email{hujinniu@nankai.edu.cn}
	\affiliation{School of Physics, Nankai University, Tianjin 300071,  China}
	\author{Ying Zhang}
	\affiliation{Department of Physics, Faculty of Science, Tianjin University, Tianjin 300072, China}
	
	\author{Hong Shen}
	\affiliation{School of Physics, Nankai University, Tianjin 300071,  China}

	\date{\today}
	\begin{abstract}
	The microscopic mechanisms of the symmetry energy in nuclear matter are investigated in the framework of the relativistic Brueckner-Hartree-Fock (RBHF) model with a high-precision realistic nuclear potential, pvCDBonn A. The kinetic energy and potential contributions to symmetry energy are decomposed. They are explicitly expressed by the nucleon self-energies, which are obtained through projecting the $G$-matrices from the RBHF model into the terms of Lorentz covariants. The nuclear medium effects on the nucleon self-energy and nucleon-nucleon interaction in symmetry energy are discussed by comparing the results from the RBHF model and those from Hartree-Fock and relativistic Hartree-Fock models. It is found that the nucleon self-energy including the nuclear medium effect on the single-nucleon wave function provides a largely positive contribution to the symmetry energy, while {the nuclear medium effect on the nucleon-nucleon interaction, i.e., the effective $G$-matrices generates the negative contribution}. The tensor force plays an essential role in the symmetry energy around the  density. The scalar and vector covariant amplitudes of nucleon-nucleon interaction dominate the potential component of the symmetry energy.  Furthermore, the isoscalar and isovector terms in the optical potential are extracted from the RBHF model. The isoscalar part is consistent with the results from the analysis of global optical potential, while the isovector one has obvious differences at higher incident energy due to the relativistic effect.  
	\end{abstract}
	
	\pacs{21.10.Dr,  21.60.Jz,  21.80.+a}
	
	\keywords{RBHF model, Nuclear matter, Symmetry energy}
	
	\maketitle

	\section{Introduction}

The nuclear symmetry energy is originally generated by the isospin degree of freedom of nucleon and the Pauli principle, which plays a very important role in the neutron-rich systems, such as the nuclei close to the neutron-drip line and the compact star in the universe~\cite{baran05,steiner05,li08,oertel17,li19}. Recently, a lot of observables about these extremely isospin asymmetry objects have been obtained from nuclear and astronomical facilities. The symmetry energy at nuclear saturation density, $E_\mathrm{sym}(n_0)$ is well constrained from the terrestrial experiments, such as the global nuclear masses and excitation energies in the nuclide chart, nuclear resonances, heavy-ion collisions, and so on~\cite{danielewicz03,danielewicz14,tsang09,tsang12,russotto16,li05,tsang19}. Its recent constraint value is $E_\mathrm{sym}(n_0)=31.6\pm2.7$ MeV through comprehensively estimating these data~\cite{li19}.


In the aspect of theoretical investigations, the symmetry energy is determined by the isospin-dependent terms in nucleon-nucleon interaction and the density of nucleons. The effective nuclear potentials based on the density functional theories are almost fixed by reproducing the ground-state properties of finite nuclei and empirical saturation properties of infinite nuclear matter, both of which are around the nuclear saturation density, $n_0$. When these nuclear many-body methods, such as  Skyrme-Hartree-Fock (SHF) model~\cite{vautherin72,stone07},  relativistic mean-field (RMF) model~\cite{walecka74,ring96,meng06, meng16}, relativistic point-coupling model~\cite{nikolaus92,zhao10}, and relativistic Hartree-Fock (RHF) model~\cite{bouyssy87,long07,meng16} are extrapolated to the high-density region, the symmetry energy presents distinct ambiguity due to the nonlinear density-dependent terms~\cite{chen05,chen12,dutra14}. Meanwhile, the {\it ab initio} methods, such as variational chain method~\cite{akmal98}, many-body perturbation method~\cite{holt17}, Brueckner-Hartree-Fock (BHF) model~\cite{li06}, relativistic Brueckner-Hartree-Fock (RBHF) model~\cite{huber95,alonso03,sammarruca12,wang20}, and so on can substantially reduce such uncertainty in symmetry energy with realistic nuclear potentials. Meanwhile, the symmetry energy and its slope were also studied in detail through the transport model, optical potential model, Glauber model, and the polynomial parameterization model with various experimental constraints~\cite{li02,li13,aumann17,zhang19}.

Moreover, many recent works attempted to explore the various microscopic mechanisms of the symmetry energy. It was found that the symmetry energy and its slope at nuclear saturation density are mainly contributed from the tensor terms of nuclear potential~\cite{vidana11,carbone14}. The three-body force generates a strong repulsive component to the symmetry energy at high densities~\cite{zuo14,goudarzi18}. The short-range correlations due to the strong repulsion core of nuclear force may also influence the kinetic and potential constitutions of the symmetry energy~\cite{rios14,hen15,cai16}. The symmetry energy can be decomposed into the nucleon self-energies with scalar and vector forms based on the Hugenholtz–Van Hove (HvH) theorem~\cite{hugenholtz58,cai12}. {The roles of meson's Fock terms in symmetry energy were also discussed in the RHF model~\cite{zhao15,sun16,liu18,miyatsu20}. When the density-dependent RHF parameter sets, which were produced by the ground-state properties of finite nuclei, were used~\cite{zhao15,sun16,liu18}, it was found that the Fock term can enhance the symmetry energy compared to the results from the RMF model.  Otherwise, when the parameter sets were obtained by fitting the nuclear saturation properties, Miyatsu et al. concluded that the RHF model suppressed the symmetry energy and its slope at high-density regions~\cite{miyatsu20}.}

{In the past few years, several important progresses on the RBHF method were achieved.  The full RBHF equations were solved for finite nuclei in a Dirac-Woods-Saxon basis and no free parameters were introduced to calculate the ground-state properties of finite nuclei and neutron drops~\cite{shen16,shen17,shen18a,shen18b,shen20}. The nuclear matter and neutron star were investigated in RBHF model without the average momentum approximation~\cite{tong18,tong20}. Furthermore, the negative-energy states were included in the Dirac space to reduce the uncertainties of single-particle potential of nuclear matter~\cite{wang21}. }

In this work, we would like to discuss the microscopic mechanism of symmetry energy from the opinions of nuclear medium effects and Lorentz covariant amplitudes of nuclear potential by comparing the results from Hartree-Fock (HF), RHF, and RBHF models. The high-precision nucleon-nucleon interaction, pvCDBonn potential will be adopted to decrease the model parameter dependence. Finally, the lowest-order isoscalar and isovector components of optical potential will be extracted from the RBHF model.

\section{The symmetry energy in Relativistic Brueckner-Hartree-Fock model}\label{sec-rbhf}
The relativistic dynamics of a nucleon in the infinite nuclear matter is described by the following Dirac equation~\cite{wang20}
\begin{equation}
	(\gamma^\mu k_\mu - M_\tau - \Sigma_\tau) u(\mathbf{k},s) = 0,
\end{equation}		
where $k^\mu$ represents the four-momentum of nucleon consisting of energy and momentum. $\tau$ and $s$ indicate its isospin and spin degrees of freedom, respectively. Because of the translation and rotation invariances of nuclear matter, the nucleon self-energy, $\Sigma_\tau$ is written as
\begin{equation}\label{se}
	\Sigma_\tau = \Sigma_\tau^\mathrm{s} - \gamma^0\Sigma_\tau^\mathrm{0}
	+\mathbf{k}\cdot \bm{\gamma} \Sigma_\tau^\mathrm{v}.
\end{equation}

Conventionally, the effective nucleon mass and momentum are defined by,
\begin{equation}\label{effME}
	M^*_\tau = { M_\tau+\Sigma_\tau^\mathrm{s} \over 1
		+\Sigma_\tau^\mathrm{v}}, \quad 
	{k_\tau^*}^\mu =  { k^\mu_\tau+\Sigma_\tau^\mu \over 1
		+\Sigma_\tau^\mathrm{v}}.
\end{equation}
Therefore, the Dirac equation in nuclear medium can be rewritten as 
\begin{equation}\label{DiracEq}
	(\bm{\alpha}\cdot\mathbf{k}+\beta M_\tau^*)	u_\tau(\mathbf{k},s) 
	= E_\tau^*(\mathbf{k}) u_\tau(\mathbf{k},s),
\end{equation}	
where $E_\tau^*(\mathbf{k})=\sqrt{\mathbf{k}^2 + M_\tau^{*2}}$ is the effective single-nucleon energy. 
The solution of above Dirac equation is a plane wave and is expressed as a spinor form with a spin wave function, $\chi_s$,
\begin{eqnarray}\label{spinor}
	u_\tau(\mathbf{k},s) = \sqrt{E^*_\tau + M^*_\tau \over 
		2M^*_\tau }\begin{pmatrix}
		1 \\
		{\bm{\sigma}\cdot \mathbf{k} \over E_\tau^* + M_\tau^*}\\
	\end{pmatrix} \chi_s.
\end{eqnarray}

The magnitudes of nucleon self-energy are determined by the nucleon-nucleon interaction in nuclear medium, $G_{\tau\tau'}$ within RBHF model, which is obtained by solving the relativistic Bethe-Brueckner-Goldstone equation~\cite{wang20,horowitz87,boelting99}. There are three available schemes to extract the nucleon self-energy from $G$-matrix in RBHF model. (1) The first one is assuming that the momentum dependence of self-energy is very weak and considering the $\Sigma$ as a constant at a fixed density. The scalar and vector components can be fitted through the single nucleon potential~\cite{brockmann90},
\begin{equation}\label{brock}
U_\tau(k)=\frac{M^*_\tau}{E^*_\tau}\Sigma_{\tau}^\mathrm{s} +\Sigma_{\tau}^\mathrm{v} ,
\end{equation}
where $\Sigma_{\tau}^\mathrm{s} $ and $\Sigma_{\tau}^\mathrm{v}$ are assumed as constants. 

{(2) It was recently extended to the second scheme, where the negative states in the Dirac space were included to reduce the uncertainties of single-particle potential. Therefore, it is a unique way to determine the nucleon self-energy and avoids the approximations in the previous method~\cite{wang21},}
 \begin{eqnarray}\label{brock}
 	&&U^{++}_\tau(k)=\frac{M^*_\tau}{E^*_\tau}\Sigma_{\tau}^\mathrm{s}(k) +\Sigma_{\tau}^\mathrm{v}(k),\\\nonumber
 	&&U^{-+}_\tau(k)=\frac{k^*_\tau}{E^*_\tau}\Sigma_{\tau}^\mathrm{v}(k),\\\nonumber
 	&&U^{--}_\tau(k)=-\frac{M^*_\tau}{E^*_\tau}\Sigma_{\tau}^\mathrm{s}(k) +\Sigma_{\tau}^\mathrm{v}(k)
 \end{eqnarray}
 where the symbol $\pm$ denote the positive  energy states and negative energy states, respectively. 

 (3) The last choice is the projection technique method using the Lorentz structure of $G$-matrix, which can keep the momentum dependence of the nucleon self-energy~\cite{horowitz87,boelting99}. In the present framework, the wave function is restricted to the positive energy states so that the ambiguity of one-pion-exchange potential is produced in the projection process, since the pseudoscalar (ps) and  pseudovector (pv) terms cannot be clearly distinguished at on-shell scattering. Hence, a subtracted representation scheme was proposed to solve such problem~\cite{dalen04,dalen07},
\begin{equation}\label{t-decmp}
	G = V^\pi_\mathrm{pv} +  V^{\omega+\rho+\sigma}_\mathrm{ps} + 
	\Delta G_\mathrm{ps}
\end{equation}

In ps representation, the $G$-matrix is separated into five covariant Lorentz amplitudes, scalar ($F^\mathrm{S}$), vector  ($F^\mathrm{V}$), tensor  ($F^\mathrm{T}$), axial-vector ($F^\mathrm{A}$), and pseudoscalar ($F^\mathrm{P}$). The nucleon self-energy is evaluated via \cite{boelting99,dalen04,dalen07},
\begin{equation}\label{psse}
	\Sigma_{\tau\tau'} = 
	\int^{|\mathbf{p}|\leqslant k_F^{\tau'}}
	{\mathrm{d}^3 \mathbf{p}\over (2\pi)^3}
	{M^*_{\tau'}F^\mathrm{S}_{\tau\tau'}+\slash\!\!\!{p^*_{\tau'}}   
		F^\mathrm{V}_{\tau\tau'}\over E^*_{\tau'}(\mathbf{p})},
\end{equation}
with $k_F^{\tau'}$ signifying the Fermi momentum for proton or neutron. Here, the antisymmetrized helicity matrix elements of $F$ are taken into account, therefore, the contributions from tensor, axial-vector, and pseudoscalar components are canceled with each other. While, in complete pv representation, the covariant Lorentz amplitudes should be expressed as interchanged Fermi covariants, and use pseudovector to replace the pseudoscalar, $g^\mathrm{S,\widetilde{S},A, PV,\widetilde{PV}}$, where $\mathrm{\widetilde{PV}}=\mathrm{PV}\mathrm{\widetilde{S} }$. The operator $\mathrm{\widetilde{S}}$ can exchange the Dirac indices of two particles in the Lorentz amplitude. The self-energy in pv representation can be calculated, 
\begin{equation}\label{pvse} 
	\begin{aligned} 
		\Sigma_{\tau\tau'}(|\mathbf{k}|) &= 
		\int
		{\mathrm{d}^3 \mathbf{p}\over (2\pi)^3}{1\over 4E^*_{\tau'}(\mathbf{p})}
		\left\{ (\slash\!\!\! k^*-\slash\!\!\! p^*)
		{2q^{*\mu}(k^*_\mu-p^*_\mu)\over (M^*_\tau + M^*_{\tau'})^2}
		g_{\tau\tau'}^{\widetilde{\mathrm{PV}}} \right. \\
		& + M^*_{\tau'} \left[ 4g_{\tau\tau'}^\mathrm{S}-
		g_{\tau\tau'}^\mathrm{\widetilde{S}} + 4g_{\tau\tau'}^\mathrm{A}
		-{(k^{*\mu}-p^{*\mu})^2 \over (M^*_\tau + M^*_{\tau'})^2}
		g_{\tau\tau'}^\mathrm{\widetilde{PV}}  \right]\\
		& + \slash\!\!\! p^* \left. \left[-g_{\tau\tau'}^\mathrm{S} 
		+2g_{\tau\tau'}^\mathrm{A}  -
		{(k^{* \mu}-p^{*\mu})^2 \over (M^*_\tau + M^*_{\tau'})^2}
		g_{\tau\tau'}^\mathrm{\widetilde{PV}}\right]\right\}.
	\end{aligned}
\end{equation}
The binding energy per nucleon in nuclear matter, $E_\mathrm{B}/A$ can be written as a function of the nucleon number density $n$ and asymmetry factor $\alpha=(n_n-n_p)/(n_n+n_p)$~\cite{wang20}. At zero temperature, according to the Hugenholtz-Van Hove (HvH) theorem~\cite{hugenholtz58}, the Fermi energy of nucleon in a thermodynamics consistent system is related to the
energy density by 
\begin{equation}\label{hvh}
	E_{\tau F} = {\partial (n_\tau E_\mathrm{B}/A) \over \partial n_\tau} 
	+ M_\tau. 
\end{equation}

{The HF and RHF models have similar theoretical frameworks as the RBHF model with the relativistic nucleon-nucleon interaction. In the HF model, the interaction between nucleons is adopted as the realistic nucleon-nucleon potential and the nucleon mass is regarded as the free nucleon mass. On the other hand, the nucleon mass and single-particle energy in the RHF model should include the nuclear medium effect through the nucleon wave function, i.e. the nucleon self-energies shown in Eqs.~\eqref{psse} and \eqref{pvse} with realistic nucleon-nucleon potential, which will be replaced by the effective $G$-matrices in the RBHF model. }

{Actually, the HvH theorem is largely violated in the lowest-order BHF approximation~\cite{zuo99} and in the RBHF model treated by the projected scheme with complete pv representation, while the violated effect is very weak by using the complete ps representation~\cite{luo06}. In present framework, the subtracted representation is used, where only the interaction of pion is projected to pv representation. Therefore, the HvH theorem should be approximately kept now and we can use it to study the Lorentz components of symmetry energy.}

The binding energy per nucleon at a fixed density, $n$ can be expanded with respect to the asymmetry factor $\alpha$,
\begin{equation}\label{expand}
	{E_\mathrm{B}\over A}(n,\alpha) =E_0(n) + \alpha ^2 E_\mathrm{sym}(n) +\cdots,
\end{equation}   
Therefore, the Fermi energy can be expressed as,
\begin{equation}\label{ef}
	\begin{aligned}
		E_{\tau F} & = M_\tau+{\partial (n E_0)\over \partial n} 
		+ 2\tau_3\alpha E_\mathrm{sym} \\
		&+\alpha^2 \left[{\partial (nE_\mathrm{sym})\over \partial n}-2E_\mathrm{sym}\right]+ \mathcal{O}(\alpha^3),\\
	\end{aligned}
\end{equation}
and the nuclear symmetry energy is able to connect to the Fermi energy as,
\begin{equation}\label{def-sym}
	E_\mathrm{sym} = {1\over 4}\left.{\partial \Delta E_F\over \partial \alpha}\right|_{\alpha =0}.
\end{equation}
where, $\Delta E_F = E_{nF}-E_{pF}$ and $\tau_3=\pm 1$ for neutron and proton, respectively.

In RBHF model, the single particle Fermi energy is $E_{\tau}(\mathbf{k}) = (1+\Sigma_\tau^\mathrm{v})E^*_\tau(\mathbf{k})
-\Sigma_\tau^0$,  which consists of the scalar and vector self-energies. Therefore the symmetry energy can be analytically expressed by the self-energy components:
\begin{equation}\label{esym-decmp}
	\begin{aligned}
		E_\mathrm{sym} & = E_\mathrm{sym}^\mathrm{kin} + E_\mathrm{sym}^\mathrm{pot},\\ E_\mathrm{sym}^\mathrm{kin} & = {k_F^2 \over 4}  \left({1+\Sigma_n^\mathrm{v}\over3E_{nF}^*}
		+{1+\Sigma_p^\mathrm{v}\over3E_{pF}^*}\right)_{\alpha=0}, \\
		E_\mathrm{sym}^\mathrm{pot} & = {1\over 4}\left[
		{M_n^*\over E_{nF}^*}{\partial \Sigma_n^\mathrm{s}\over 
			\partial \alpha}
		-{M_p^*\over E_{pF}^*}{\partial \Sigma_p^\mathrm{s}\over
			\partial \alpha}
		-{\partial  (\Sigma_n^0-\Sigma_p^0)\over \partial \alpha}\right.
		\\ 
		&\left. +k_F^2 \left({1\over E_{nF}^*}
		{\partial \Sigma_n^\mathrm{v}\over \partial \alpha}-
		{1\over E_{pF}^*}{\partial \Sigma_p^\mathrm{v}\over \partial
			\alpha}\right)\right]_{\alpha=0},
	\end{aligned}
\end{equation}
with the average Fermi momentum $k_F = \left({3\pi^2 n/2}\right)^{1\over 3}$.

The optical nucleon potential is very essential for nucleon-nucleus scattering calculations and has strong isospin dependence. It is obtained in RBHF model by reducing the Dirac equation~\eqref{DiracEq} to a Schr\"{o}dinger-equivalent equation, $H ={\mathbf{k}^2\over 2M_\tau} + U_\tau^\mathrm{op}$ ~\cite{jaminon89,dalen05}
\begin{equation}\label{oppot}
	\begin{aligned}
		U^\mathrm{op}_\tau &= \Sigma_\tau^\mathrm{s}- 
		{E_\tau\over M_\tau}\Sigma_\tau^\mathrm{0}+{\mathbf{k}^2\over M_\tau}
		\Sigma_\tau^\mathrm{v}+{{\Sigma_\tau^\mathrm{s}}^2
			-{\Sigma_\tau^\mathrm{0}}^2
			+ \mathbf{k}^2{\Sigma_\tau^\mathrm{v}}^2\over 2M_\tau},
	\end{aligned}
\end{equation}
which is expanded with respect to the asymmetry factor $\alpha$, $U^\mathrm{op}_\tau = U^\mathrm{op}_0 +\tau_3 \alpha
U^\mathrm{op}_\mathrm{sym}+\cdots$. The first term corresponds to the isoscalar potential, and the second one is isospin dependent, so-called the
Lane potential~\cite{lane62}, which can be extracted from the nucleon-nucleus scattering data.
\begin{equation}\label{uop01}
	U_\mathrm{0}^\mathrm{op} = 
	{U_n^\mathrm{op}+U_p^\mathrm{op}\over 2}, \quad 
	U_\mathrm{sym}^\mathrm{op} = 
	{U_n^\mathrm{op}-U_p^\mathrm{op}\over 2\alpha}.
\end{equation}

\section{The numerical results and discussions}
In Table~\ref{table.1}, the nuclear saturation properties of symmetric nuclear matter, i.e. saturation density, $\rho_0$, the binding energy per nucleon, $E/A$, and incompressibility, $K$ are given with different schemes in RBHF model, which were mentioned  in section {\ref{sec-rbhf}}. The Bonn A, B, C potentials are chosen as the realistic nucleon-nucleon interactions. The momentum-dependence of self-energy is neglected in the scheme (1). The negative energy states is considered in the scheme (2). In the scheme (3), the $G$-matrix are project to five covariant Lorentz structures. These saturation properties from scheme (1) and scheme (2) are very similar since the components of self energies are obtained from the single-particle potential in these two methods, while the binding energy per nucleon from the scheme (3) have a slight differences from those from scheme (1) and (2), since some part of $G$-matrix are projected as ps amplitude, which can generate more attractive contributions comparing to the pv amplitude.

\begin{table*}[htb]
	\centering
	\scalebox{0.85}{
		\begin{tabular}{ccccc}
			\hline\hline
			   Scheme     &   Potential      &$\rho_0[\rm fm^{-3}]$    &$E/A [\rm MeV]$ &$K[\rm MeV]$   \\
			\hline                          
			         &Bonn A  &0.180     &-15.38   &286    \\
	(1) \cite{tong18}   &Bonn B  &0.164     &-13.44   &222    \\
			        &Bonn C  &0.149    &-12.12    &176    \\
			        \hline
		            &Bonn A  &0.188     &-15.40   &258   \\
	       (2)\cite{wang21}   &Bonn B  &0.164     &-13.36   &206    \\
		            &Bonn C  &0.144    &-12.09   &150    \\
		            \hline
		            &Bonn A  &0.179    &-16.18   &250    \\
		 (3)     &Bonn B  &0.163    &-14.63  &200    \\
		            &Bonn C  &0.149    &-13.68    &170    \\
			\hline\hline
	\end{tabular}}
\caption{ The nuclear saturation properties of symmetric nuclear matter from different schemes in RBHF model.}\label{table.1}
\end{table*}

In the following calculations, pvCDBonn A potential will be used as the input realistic nuclear force. It is a high-precision charge-dependent potential including the explicit charge symmetry breaking (CSB) and charge independence breaking (CIB) effects and has a relatively small tensor force component, $D$-state probability of deuteron $P_D=4.2\%$~\cite{wang19}, which describes the properties of symmetric nuclear matter and neutron star very well in RBHF model~\cite{wang20}. 

To study the nuclear medium effect on the symmetry energy, three types of calculations will be performed:
(1) The binding energy per nucleon and symmetry energy of nuclear matter will be generated directly with pvCDBonn A potential by Hartree-Fock (HF) model, where the nucleon wave function is a non-relativistic plane wave.
(2) The nuclear medium effect will be taken into account in the nucleon propagator. The mass and single-particle energy of nucleon are dressed by nucleon self-energies self-consistently in the mean-field method, i. e., RHF model.
(3) The Bethe-Goldstone  equation will be solved to include the medium effect both on the nuclear potential and wave function with subtracted scheme of project method~\cite{dalen04}, based on the RHF model, and achieve the RBHF calculation.

\begin{figure}[htb]
	\centering
	\includegraphics[width=0.5\linewidth]{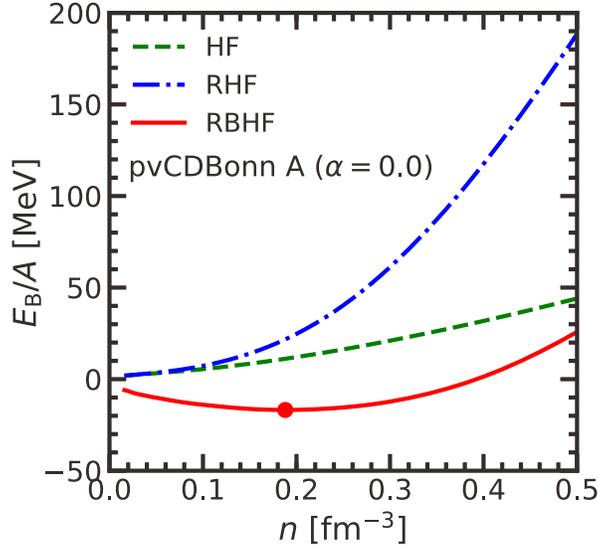}
	\caption{ The binding energy per nucleon in symmetric nuclear matter
		calculated with pvCDBonn A potential in the frameworks of HF, RHF, and RBHF models
		The point RBHF curve indicates the saturation point of 
		pvCDBonn A ($n_\mathrm{sat}=0.19$ fm$^{-3}$,
		$E_\mathrm{B}/A(n_\mathrm{sat})=-16.89$ MeV).}
	\label{fig:eb}
\end{figure}

The results of binding energy per nucleon in symmetric nuclear matter from three types of calculations are presented in	Fig.~\ref{fig:eb}. In the mean-field approximation, the strong repulsion of pvCDBonn potential at a short-range distance cannot be properly handled. Hence, there is no  bound state in symmetry nuclear matter, where the $E/A>0$ for all densities. Then, the nuclear medium effect is introduced into the nucleon propagator through the Dyson equation. The self-energies appear in the denominator of the nucleon propagator and influence the nucleon mass and single-particle energy. They can be solved self-consistently in a relativistic framework. The equation of state (EOS) of symmetric nuclear matter from the RHF model is obvious stiffer than that from the HF model, especially in the high-density region. It is because that the nucleon-antinucleon excitation through exchanging the scalar mesons, i.e., $Z$-diagram can generate a very strong positive contribution to the binding energy.  In the RBHF model, the short-range repulsion is removed by summing all ladder diagram of nucleon-nucleon scattering in nuclear medium with Bethe-Goldstone equation. As a result, the binding energy at the low-density region becomes negative. Its magnitude at saturation density, $E_\mathrm{B,sat}/A=-16.69$ MeV at $n_\mathrm{sat}=0.19$ fm$^{-3}$  almost reproduces the empirical saturation properties of symmetric nuclear matter.

\begin{figure}[htb]
	\centering
	\includegraphics[width=0.5\linewidth]{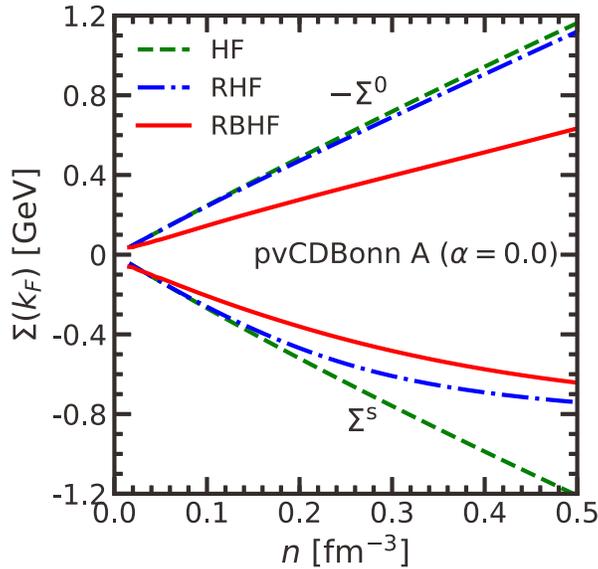}
	\caption{The self-energies as functions of density obtained from HF, RHF, and RBHF models.}
	\label{fig:se}
\end{figure}

In Fig.~\eqref{fig:se}, the scalar and time-component of vector self-energies at Fermi surface $k_F$, $\Sigma^\mathrm{s}$ and $\Sigma^0$ in symmetric nuclear matter are presented from HF, RHF, and RBHF methods, respectively. At HF level, we directly calculated them with the free nucleon propagator in Eqs.~\eqref{psse} and \eqref{pvse}, while the interacting propagator was adopted in RHF and RBHF models. The $\Sigma^0$ in HF and RHF models are almost the same since it is only dependent on the covariant amplitudes, $F^\mathrm{V}$ and $g^\mathrm{i}$, which have a few differences in HF and RHF models due to the effective mass and single-particle energy. Meanwhile, scalar self-energies from the RHF model are significantly larger than those from HF model due to the medium effect on the effective nucleon mass. Therefore, the EOS of symmetric nuclear matter from the RHF model is much stiffer. Furthermore, the magnitudes of $\Sigma^\mathrm{s}$ and  $\Sigma^0$ from the RBHF model are both smaller than those from the previous two models. The nuclear medium effect renormalizes the realistic nuclear potential to an effective one, whose covariant amplitudes are changed completely  comparing to those in HF and RHF models. The reduction of vector self-energies leads to the bound states of the nuclear many-body systems at low-density regions.

\begin{figure}[htb]
	\centering
	\includegraphics[width=0.5\linewidth]{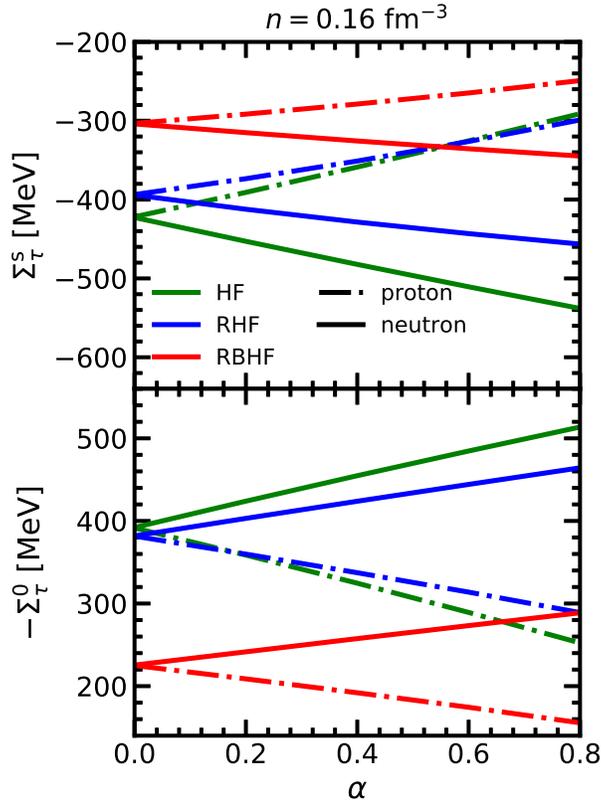}
	\caption{The self-energies as functions of asymmetry factor at empirical nuclear saturation density obtained from HF, RHF, and RBHF models.}
	\label{fig:iso}
\end{figure}

{The isospin dependence of self-energies at empirical nuclear saturation density, $n_0=0.16$ fm$^{-3}$ from the HF, RHF, and RBHF model is given in Fig.~\eqref{fig:iso}. With the asymmetry factor, $\alpha$ increasing, the differences of the self-energies from neutron and proton become larger and the scalar self-energy of the neutron is lower than that of the proton, which is consistent with the conclusions from the RHF model. Furthermore, the splittings between proton and neutron self-energies from the HF model are the largest among the three models, with the nuclear medium effect is included, the splitting in the RBHF model at $\alpha=0.8$ is just half of that in the HF model. Furthermore, the self-energies in these methods almost linearly increase with $\alpha$.}

\begin{figure}[htb]
	\centering
	\includegraphics[width=0.5\linewidth]{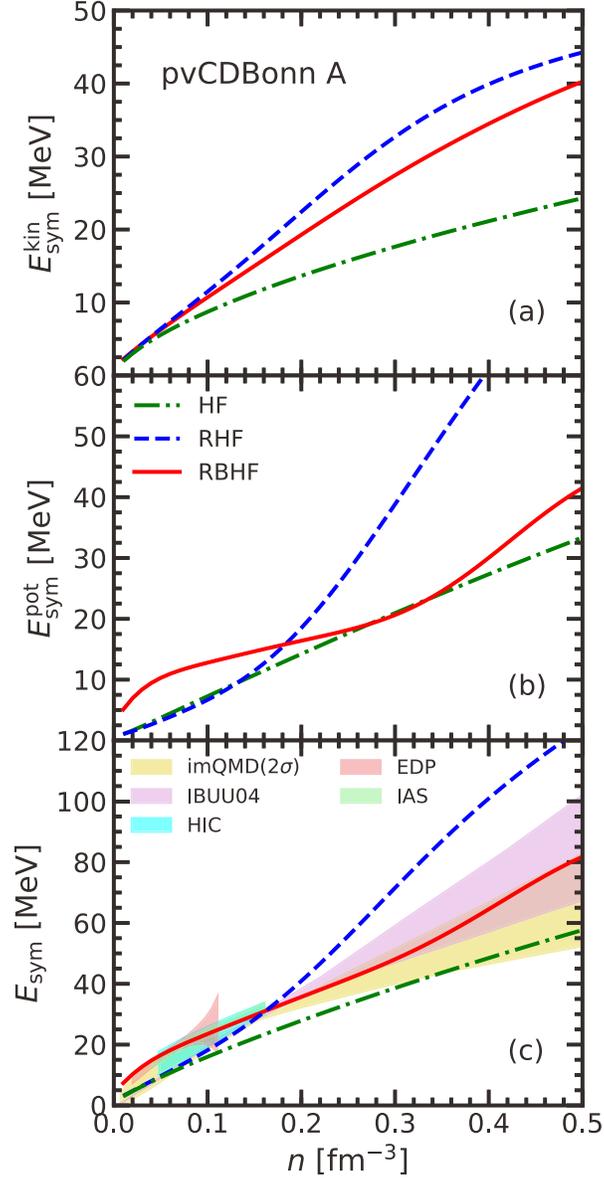}
	\caption{The symmetry energy and its components obtained in the three models.
		The shaded regions in panel (c) are constraints from various
		sources, explained in the context.}
	\label{fig:esymdcmp}
\end{figure}

The characters of nucleon self-energies will largely influence the behaviors of the nuclear symmetry energy. In Fig.~\eqref{fig:esymdcmp}, the symmetry energy, and its kinetic energy, and potential components from HF, RHF, and RBHF models are shown respectively.  In panel (a), the kinetic-energy contributions of symmetry energy $E^\mathrm{kin}_\mathrm{sym}$ from three models are plotted. All of them increase with nuclear density. The HF results correspond to the free Fermi gas. The effective nuclear mass in RHF will reduce the effective single-particle energy, while the vector self-energies are also identical in the HF and RHF models. Therefore, the $E^\mathrm{kin}_\mathrm{sym}$ from RHF is much larger than that generated by the HF model through Eq.~\eqref{esym-decmp}.  Although the scalar self-energy from the RBHF model is the largest among the three results, its vector one also decreases due to the medium effect on the potential. Therefore, its $E^\mathrm{kin}_\mathrm{sym}$ is less than that from the RHF model. 

The potential contributions of symmetry energy, $E^\mathrm{pot}_\mathrm{sym}$ are given in the panel (b) of Fig.~\eqref{fig:esymdcmp}. The curve from HF model increases as a linear relation with the density, which is determined by the similar behaviors of the scalar and vector self-energies as shown in Fig.~\eqref{fig:se}. The potential contribution of symmetry energy from the RHF model below the saturation density is almost identical to that from the HF level, where the strengths of scalar fields are still small. It grows rapidly in the high-density region and shows a strong relativistic effect. Meanwhile, the result generated from the RBHF model shows a completely different tendency comparing to the other two models. At low density, the $E^\mathrm{pot}_\mathrm{sym}$ in the RBHF model is much larger than those from HF and RHF methods, since the tensor force plays a very important role below the saturation density in symmetric nuclear matter~\cite{wang20}, which cannot be treated at the mean-field level but was taken into account in RBHF model. With density increasing, the part of the tensor force is weakened, while nucleons are closer to each other and the short-range correlation becomes significant, which will be shown in detail later. This medium effect will suppress the potential contribution of the symmetry energy.

The total symmetry energies of nuclear matter, $E_\mathrm{sym}$ as functions of density are plotted in the panel (c) of Fig.~\eqref{fig:esymdcmp}. Its value at saturation density is $E_\mathrm{sym}(n_\mathrm{sat})=34.48$ MeV. At the same time, the recent constraints from various experiments on  $E_\mathrm{sym}$, such as the heavy-ion collision (HIC)~\cite{tsang12}, electric dipole polarizability (EDP) of $^{208}$Pb~\cite{zhang15}, isobaric analog states joint with neutron skin thickness~\cite{danielewicz14} (IAS), the improved quantum molecular dynamics calculation  (imQMC, $2\sigma$ confidence region)~\cite{tsang09}, and transport model simulation of isospin diffusion experiment (IBUU04)~\cite{chen05}, are also given. It can be found that the symmetry energy from RBHF satisfies all these constraints in the whole density region. Due to the tensor effect, the symmetry energy in RBHF at low density can describe the data from HIC, EDP, and IAS better. Meanwhile, the RHF model provides too large symmetry energy without the high-momentum correlations. In the work of Cai and Li, it has already been pointed that the high-momentum contributions can reduce the symmetry energy~\cite{cai16}.

\begin{figure}[htb]
	\centering
	\includegraphics[width=0.5\linewidth]{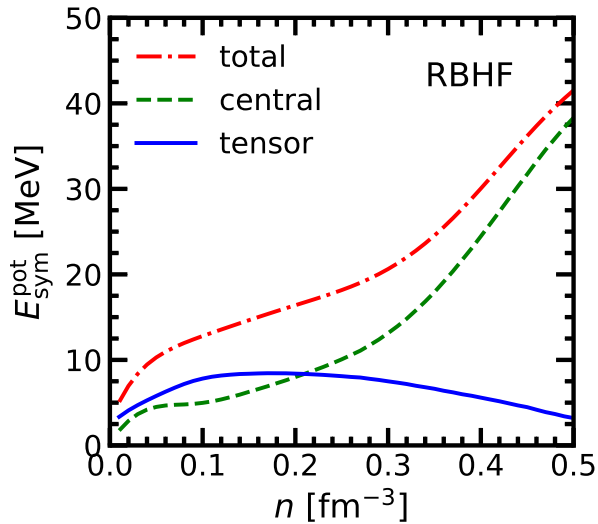}
	\caption{The contributions of tensor and central forces on the potential components of symmetry energy.}
	\label{fig:tencom}
\end{figure}

{In the RBHF model, it is very difficult to clearly distinguish the roles of various components of realistic nucleon-nucleon interaction due to the iterated process. A good approximation was proposed in Ref.~\cite{nosyk21} to discuss the tensor force contribution in the BHF model, where it was regarded as the second-order perturbation term of tensor force.  In this work, the same schemes are adopted to show the tensor force contribution to the potential component of symmetry energy in Fig.~\eqref{fig:tencom}. It can be found that the tensor force plays an essential role around the saturation density region, while it becomes weaker with density increasing. On the other hand, the tensor effect on the symmetry energy was also discussed in DDRHF model~\cite{jiang15}, where, the tensor force provided the negative contribution on symmetry energy. It may be caused by the different definitions of tensor force between their work and the present one. } 

\begin{figure}[htb]
	\centering
	\includegraphics[width=0.5\linewidth]{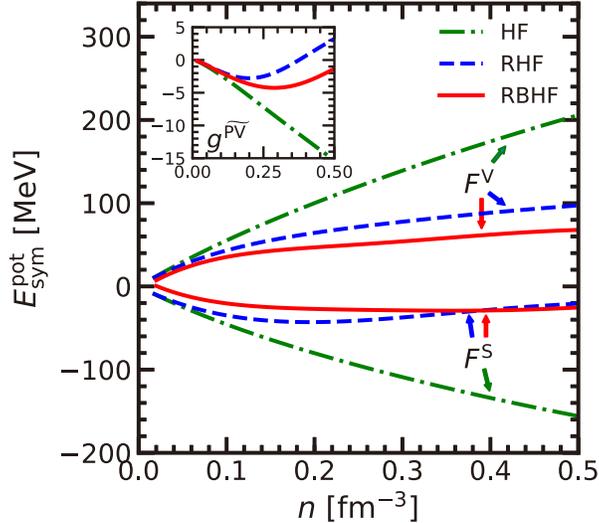}
	\caption{The covariant amplitude contributions in the potential component of symmetry energy from HF, RHF, and RHF models.}
	\label{fig:cov}
\end{figure}

The symmetry energy from the potential part also can be decomposed into the covariant Lorentz structure in the project scheme of the RBHF model. The scalar, vector, and pseudovector terms are shown in Fig.~\eqref{fig:cov}. In the present subtracted-$G$ matrix way, the one-pion exchange potential is projected to the pv representation and only has the pseudovector amplitude, $g^{\widetilde{\mathrm{PV}}}$~\cite{plohl06}. The rest part in $G$-matrix is projected to ps representation and transferred to the scalar and vector covariant amplitudes, $F^{\mathrm{S}}$ and $F^{\mathrm{V}}$ after taking the antisymmetrized helicity matrix elements. The contributions from scalar and vector components are gradually cut down from HF model to RBHF model but always dominate the $E^\mathrm{pot}_\mathrm{sym}$, while those from the  pseudovector amplitude are very small due to the Fock term and only provide few attractive contributions in the HF model and RBHF model, which is opposite to the recent results about the role of the pion in RHF model~\cite{miyatsu20}.  Here, we must emphasize that in the conventional treatment, the contact component in one-pion-exchange potential will be removed in RHF model~\cite{bouyssy87}. Therefore, the pion contribution has an opposite sign with us.

\begin{figure}[htb]
	\centering
	\includegraphics[width=0.5\linewidth]{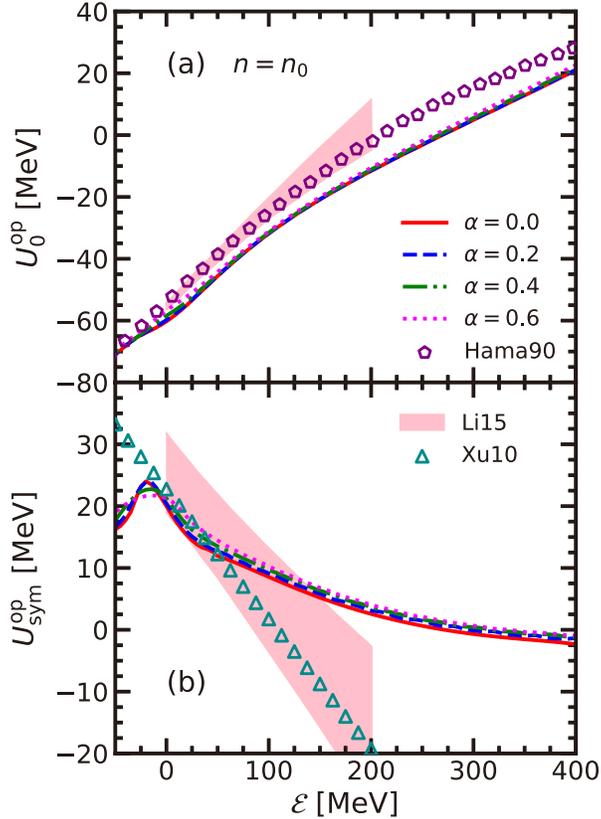}
	\caption{The isoscalar optical potential and the Lane potential,
		at $n_0$ obtained with different $\alpha$, as functions of incident energy.}
	\label{fig:Uopt01}
\end{figure}

Finally, the real part of optical potential, $U_\tau^\mathrm{op}$ is extracted from RBHF model with the Sch\"{o}dinger-equivalent potential. Its isoscalar and isovector component as functions of  incident energy $\mathcal{E} =(1+\Sigma_\tau^\mathrm{v})E_\tau^*(\mathbf{k})-\Sigma_\tau^0 - M_\tau$ at empirical saturation density, $n_0=0.16$ fm$^{-3}$ with different asymmetry factors, $\alpha$ are presented in Fig.~\ref{fig:Uopt01}. These optical potentials are almost identical with different $\alpha$.
The corresponding analysis by global Dirac optical model (Hama90)~\cite{hama90},  averaged global optical potentials (Xu10)~\cite{xu10}, and nonrelativistic optical models (Li15)~\cite{li15} is also shown to be compared. The isoscalar optical potential, $U_0^\mathrm{op}$ from the RBHF model,  monotonously increases with the incident energy and is consistent with the analysis by Hama and Li {\it{et al.}}~\cite{hama90,li15}. However, for the isovector component, $U_\mathrm{sym}^\mathrm{op}$, i.e., Lane potential, it has the completely different behavior from the RBHF model, which is obviously larger than those from the analysis with nonrelativistic optical models~\cite{xu10,li15} and slowly decreases at higher incident energies. In this region, there is not enough experimental data until now, which should be clarified in the future.

\section{Summary and outlook}\label{sec-concl}
The nuclear symmetry energy was investigated in the framework of the relativistic-Brueckner-Hartree-Fock (RBHF) model within a high-precision nuclear potential, pvCDBonn potential, which perfectly satisfies the various constraints from recent experiments. The kinetic energy and potential components in symmetry energy were decomposed into the nucleon scalar and vector self-energies, which were provided by the project scheme of RBHF models. 

By comparing the results from Hartree-Fock (HF) and relativistic Hartree-Fock (RHF) models, it is found the relativistic effect provides a very strong repulsive contribution to the symmetry energy, {while the nuclear many-body medium effect generated by the RBHF model will reduce the kinetic energy part of the symmetry energy}. The tensor force plays a significant role around ion density for the potential terms of the symmetry energy. The potential component of symmetry energy can be further separated into various Lorentz covariant amplitudes. The main contributions are generated by the scalar and vector amplitudes, whereas, the pseudovector amplitude from one-pion-exchange potential only provides a few attractions. 

The real part of nucleon optical potential was also extracted in the RBHF model and was compared to the recent analysis by global optical potential models. The isoscalar terms of optical potential from present calculations are consistent with the available analysis, however, the isovector optical potential term from the RBHF model has obvious differences and decreases with the incident energy slowly due to the relativistic effect. Therefore, the relativistic optical potential analysis in the nucleon-nucleus scattering will be done in the future.

\section*{Acknowledgments}
This work was supported in part by the National Natural Science Foundation of China (Grant  Nos. 11775119 and 12175109), and  the Natural Science Foundation of Tianjin.

\end{document}